# Entanglement of charge transfer, hole doping, exchange interaction and octahedron tilting angle and their influence on the conductivity of $La_{1-x}Sr_xFe_{0.75}Ni_{0.25}O_{3-\delta}$:

# A combination of x-ray spectroscopy and diffraction

Selma Erat[1,2,a], Artur Braun[1,b], Cinthia Piamonteze[3], Zhi Liu[4], Alejandro Ovalle[1],

Hansjürgen Schindler[1], Thomas Graule[1,5], Ludwig J. Gauckler[2]

[1]Laboratory for High Performance Ceramics

EMPA – Swiss Federal Laboratories for Materials Testing and Research

CH-8600 Dübendorf, Switzerland

[2]Department of Materials, Nonmetallic Inorganic Materials

ETH Zurich – Swiss Federal Institute of Technology

CH-8093 Zurich, Switzerland

[3]Swiss Light Source, Paul Scherrer Institut

CH-5232 Villigen PSI, Switzerland

[4]Advanced Light Source, Ernest Orlando Lawrence Berkeley National Laboratory

Berkeley CA 94720, USA

[5]Technische Universität Bergakademie Freiberg,

D-09596 Freiberg, Germany

**Abstract:**

Substitution of La by Sr in the 25% Ni doped charge transfer insulator LaFeO$_3$ creates structural changes that inflect the electrical conductivity caused by small polaron hopping via exchange interactions and charge transfer. The substitution forms electron holes and a structural crossover from orthorhombic to rhombohedral symmetry, and then to cubic symmetry. The structural crossover is accompanied by a crossover from Fe$^{3+}$–O$^{2-}$–Fe$^{3+}$ superexchange interaction to Fe$^{3+}$–O$^{2-}$–Fe$^{4+}$ double exchange interaction, as evidenced by a considerable increase of conductivity. These interactions and charge transfer mechanism depend on superexchange angle, which approaches 180° upon increasing Sr concentration, leading an increased overlap between the O (2p) and Fe/Ni (3d) orbitals.

[a,b]Corresponding authors: selma.erat@empa.ch and artur.braun@alumni.ethz.ch , Phone: +41 44 823 4971/4850, Fax: +41 44 823 4150.

**Introduction**

Anions in ionic compounds, such as oxygen in $ABO_3$ perovskites, play a decisive role in the electronic transport properties of these materials. For example, the oxygen mediates electron hopping between its neighboring cations, here Fe and Fe, or Fe and Ni, by exchange interactions where the oxidation state and the spin state of the metal ions determine whether hopping across oxygen can take place or not [1, 2]. Particularly, substitution in mixed valence compounds containing magnetic metal ions shows how conductivity is heavily influenced by these exchange interactions and corresponding magnetovolume effects, which goes along with changes in lattice spacing and crystallographic symmetry.

Spin related exchange interaction, also referred to as superexchange, is the antiferromagnetic coupling between two next-to-nearest neighbor positive ions through a non-magnetic anion, such as oxygen [2]. It strongly depends on the electronic and crystallographic structure such as electron occupancy, orbital configuration and geometry, respectively [3]. In the case of double exchange, the electrons move between positive ions having different d-shell occupancy via a non-magnetic anion [1].

The electronic conductivity of perovskites is generally explained in terms of small polarons which are thermally activated [4]. The electrons hop from one side to the other via B–O–B bridge which are mostly increased by overlap (strongly depends on the B–O distance and B–O–B superexchange angle [5]. A-site substitution in iron perovskites is well studied and relatively well understood. B-site substitution is less well studied. At the molecular scale, the structure of the perovskite includes the valence state of the A-site and B-site cations, and the spin state of the B-site cations, and the oxygen deficiency.

$LaFeO_3$ is a well known $ABO_3$ perovskite with orthorhombic symmetry (a=5.5647 Å, b=7.8551 Å, and c=5.5560 Å) and an antiferromagnetic insulator with a Neel temperature $T_N$=750 K [6]. When the trivalent La is substituted by the divalent Sr, which also has a larger ion radius than La, one electron hole is created at the oxygen site, and the size difference constitutes a chemical pressure [7]. This effect decreases the rhombohedral lattice distortion. The end member of the substitution, $SrFeO_3$, has cubic symmetry, which in turn increases overlap between O (2p) and Fe (3d) orbitals [8]. In response to the electron hole created, Fe oxidizes from $Fe^{3+}$ towards $Fe^{4+}$ with a parallel increase in the conductivity.

Recently, oxide perovskites such as members of $LaFe_{1-y}Ni_yO_{3-\delta}$ [9-12], and $La_{1-x}Sr_xFe_{1-y}Ni_yO_{3-\delta}$ [13, 14] have received much attention. Since they show high electronic conductivity and exhibit sufficient stability at elevated temperatures, they are considered as good cathode materials for solid oxide fuels cells. Such complex perovskites are members of strongly correlated electron system and also received much attention from the point of electronic structure. For example, Sarma et. al. [15], Kumar et. al. [16] worked on $LaFe_{1-y}Ni_yO_{3-\delta}$; Abbate et. al. [17], Chainani et. al. [18] and Wadati et. al. [19] worked on $La_{1-x}Sr_xFeO_{3-\delta}$.

We have recently presented a detailed experimental soft x-ray absorption study on $La_{1-x}Sr_xFe_{0.75}Ni_{0.25}O_{3-\delta}$ at Fe L edge supported by Ligand Field Multiplet Calculation [20] and at O K edge [21]. We showed two different mechanisms which affect electrical conductivity: d-type ($Fe^{4+}/(Fe^{4+}+Fe^{3+})$) and p-type $[e_g(\uparrow)/(t_{2g}(\downarrow) + e_g(\downarrow))]$ electron holes created on the Fe and O site, respectively. One conclusion of this study was that p-type electron holes are mainly caused by charge transfer from O (2p) to Ni (3d) rather than Fe (3d) orbitals.

In the present work we discuss the changes in the crystallographic structure upon Sr doping (i. e. symmetry, Fe/Ni–O distance, superexchange angle) which affect hopping process ($Fe^{3+}$–$O^{2-}$–$Fe^{3+}/Fe^{4+}$) with either superexchange or double exchange and charge transfer process and consequently electronic conductivity. An increase in Tolerance factor is paralleled to an increase in symmetry. We also discuss the effect of the A-site Coulomb potential on the B-site which also plays an important role on the conductivity.

**Experimental Section**

La$_{1-x}$Sr$_x$Fe$_{0.75}$Ni$_{0.25}$O$_{3-\delta}$ (LSFN) with x=0.0, 0.25, 0.50, 0.75, 1.0 and LaFeO$_3$ were prepared by conventional solid state reaction. The precursors La$_2$O$_3$ (>99.99 %), SrCO$_3$ (99.9 %), Fe$_2$O$_3$ (>99.0 %) and NiO (99.8 %) were mixed in stoichiometric proportions, calcined at 1200 °C for 4 h and then sintered at 1400 °C for 12 h with heating/cooling rate of 5 K/min. for LSFN. LaFeO$_3$ was calcined at 1200 °C for (4h + 4h) with same heating/cooling rate. X-ray powder diffractograms (XRD) were collected with a Philips X'Pert PRO-MPD diffractometer at ambient temperature (40 kV, 40 mA, Cu-K$_\alpha$ $\lambda$=1.5405 Å) in steps of 0.02° for 20° ≤ 2θ ≤ 80°. Rietveld structure refinement was performed with GSAS [22, 23].

Near edge x-ray absorption fine structure (NEXAFS) spectra at 300 K were recorded at the Advanced Light Source in Berkeley, Beamline 9.3.2, the end station of which has an operating energy range of 200-1200 eV and an energy resolution of 1/10000. The vacuum chamber base pressure was lower than 5x10$^{-10}$ Torr. Signal detection was made in total electron yield (TEY) mode. Powder samples were dispersed on conducting carbon tape and then mounted on a copper sample holder. Iron L-edge spectra were recorded from 690 to 750 eV, Oxygen K-edge spectra were recorded from 520 to 560 eV in steps of 0.1 eV.

For the conductivity measurements, the calcined powders were pressed into bars with dimensions of about 5 mm x 3 mm x 25 mm and sintered at 1400 °C for 12 h with heating/cooling rate of 5 K/min. Four Pt terminals were applied on the sintered bars using Pt paint (CL11-5100, W. C. Heraeus GmbH & Co. KG, Germany) and calcined with heating/cooling rate of 5 K/min up to 1000 °C with 45 min. dwell time at 1000 °C, and then cooled down to ambient temperature.

## Results

**Crystallographic structure**

The evolution of the crystallographic structure, as reflected by the X-ray diffractograms in Figure 1 shows that LSFN undergoes structural transformation upon substitution of La by Sr. The visually best match can be made for the end members with x=0 for orthorhombic and with x=1.0 for cubic symmetry. For the intermediate mixed members (x = 0.25, 0.50, 0.75), clear distinction between cubic and orthorhombic symmetry cannot be unambiguously made without deeper analysis of the diffraction data, which will follow in this section.

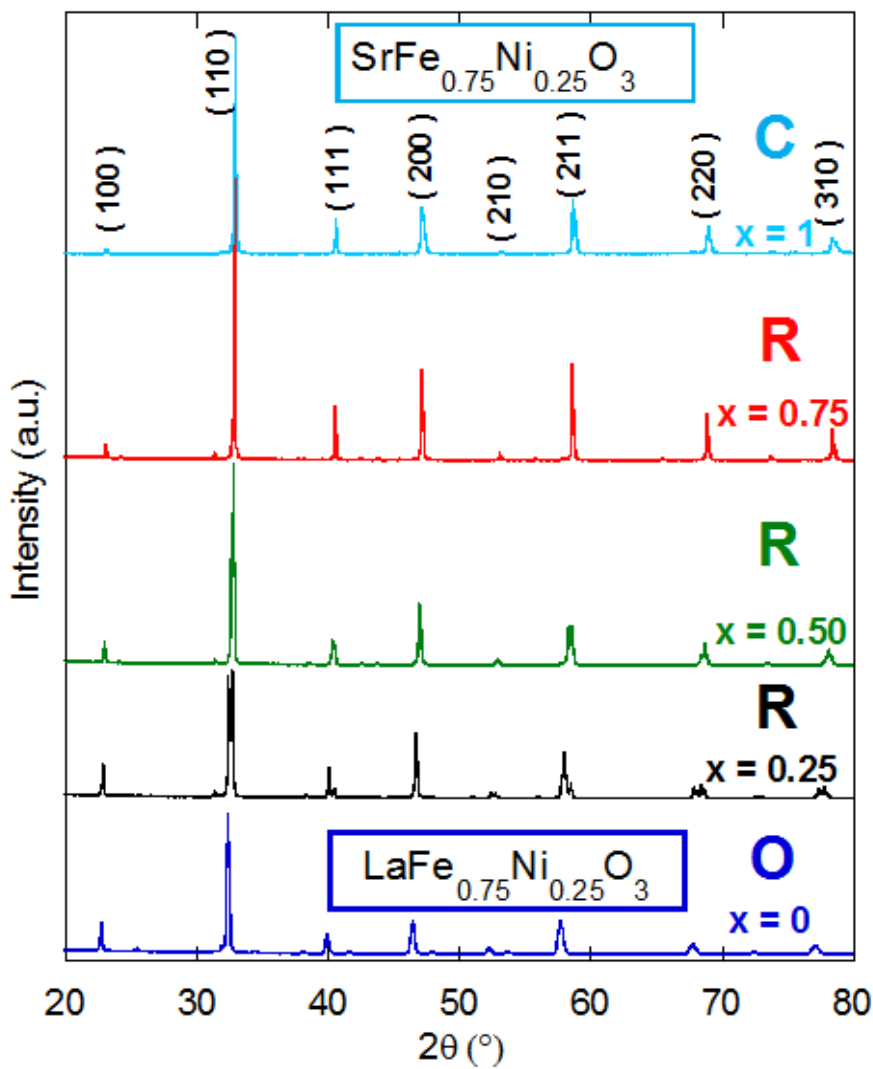

Figure 1. X-ray diffractograms for $La_{1-x}Sr_xFe_{0.75}Ni_{0.25}O_{3-\delta}$; x=0 orthorhombic (O); x=0.25, 0.50, 0.75 rhombohedral (R); 1.0 cubic phase (C). The pure perovskite phase is indexed with all Bragg reflections as the cubic phase on top of the patterns.

The Bragg reflections shift towards larger diffraction angles with increasing Sr substitution, revealing a general trend of decreasing unit cell volumes. For x=0, the (110) reflection is virtually a single peak, which however splits into a distinct double peak for x=0.25 in the rhombohedral phase. For x=0.50 and 0.75, the splitting gets diminished. The peak splitting decreases with increasing Sr content from 0.25 to 0.50. For x=0.50 the peak splitting in not so clear but still shows an asymmetric shape. For x=0.75 no clear peak splitting is observed but a small shoulder appears on the right side. The gradual decrease in the peak splitting can be considered as a second order phase transformation to higher crystal symmetry [24].

It is known that the transport and magnetic properties of perovskites are sensitive to even minute structural imbalances such as tilting of the $MO_6$ (M indicates transition metals) octahedra. In particular, parameters like bond length and bond angle determine orbital overlap and thus charge transfer and exchange interactions, with potential effect on the transport properties. In order to quantify such structural details and changes, Rietveld analysis [25] was applied. The comparison of experimental, calculated patterns and the differences are shown for x=0.0, x=0.50 and x=1.0 in Figure 2.

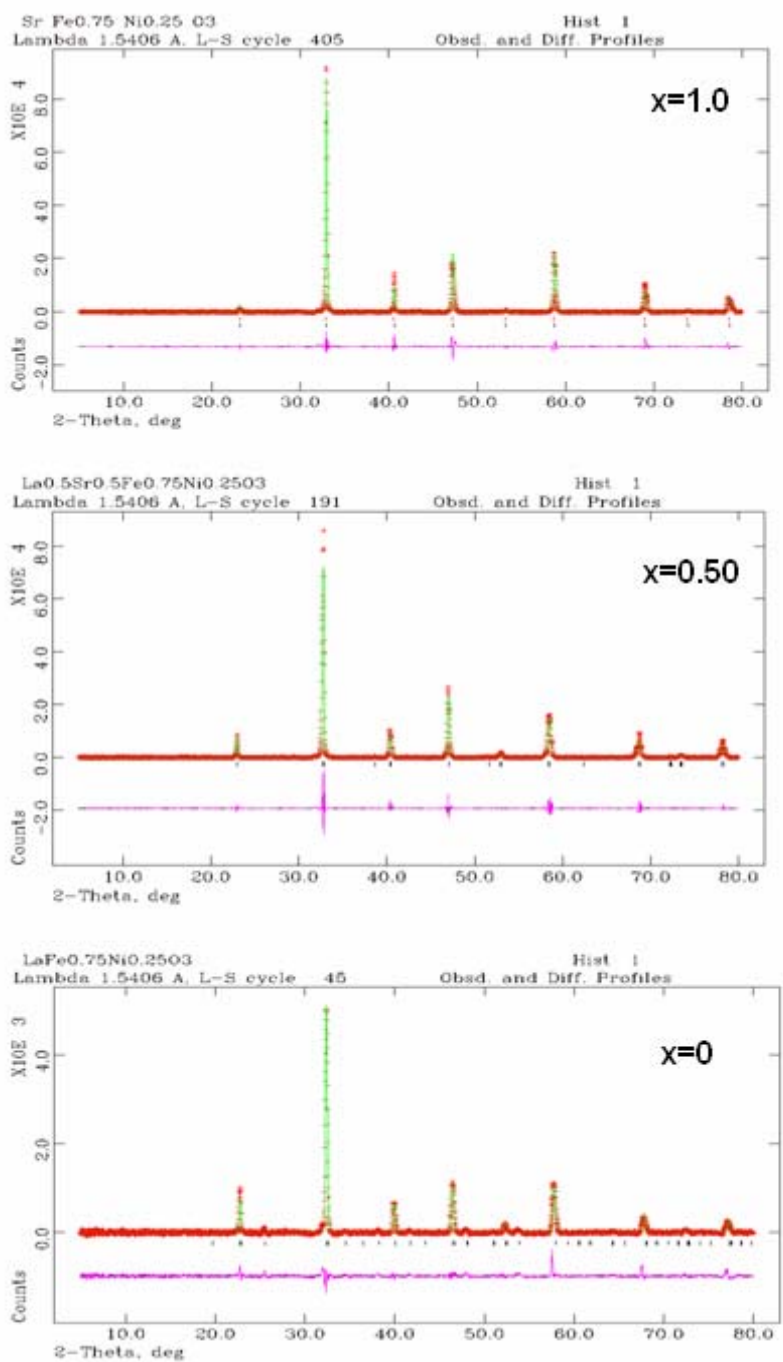

Figure 2. Rietveld refinement profile of LaFe$_{0.75}$Ni$_{0.25}$O$_{3-\delta}$ (x=0), La$_{0.5}$Sr$_{0.5}$Fe$_{0.75}$Ni$_{0.25}$O$_{3-\delta}$ (x=0.50) and SrFe$_{0.75}$Ni$_{0.25}$O$_{3-\delta}$ (x=1.0). Experimental pattern (red), calculated pattern (green), difference pattern (bottom) are shown.

Orthorhombic symmetry with space group Pbnm (62) could be confirmed for x=0. Pbnm is a pseudo-cubic space group and the unit cell parameters are related to the ideal cubic perovskites as a≈$\sqrt{2}$ $a_p$, b≈$\sqrt{2}$ $a_p$, and c≈$2a_p$. In Glazer's notation this tilting system is written as ($a^+b^-b^-$) [26]. For LSFN with x=0.25, 0.50, and 0.75 Rietveld analysis yields rhombohedral symmetry with space group R-3c (167). Rhombohedral distortion is a consequence of equivalent antiphase octahedra tilting along three crystallographic axes, i.e., a-a-a- following Glazer's notation [26]. Finally, LSFN with x=1.0 has cubic symmetry with space group Pm-3m (221). The samples with mixed phase of La and Sr show a very small peak at around 31.3° due to minor contamination by a tetragonal phase that we could match with a nickelate phase (JCPDS 01-081-2084, $La_{1.71}Sr_{0.19}NiO_{3.9}$).

The unit cell parameters (a, b, c), volume (V), interatomic distance for Fe/Ni–O, average superexchange angle and average tilting angle <ω> determined by Rietveld structure refinement are summarized in Table 1.

**Table 1**: Refined structure parameters for $La_{1-x}Sr_xFe_{0.75}Ni_{0.25}O_{3-\delta}$.

| x | Symmetry | a(Å) | b(Å) | c(Å) | V(Å³) | Fe/Ni–O (Å) | Superexchange angle (°) | Tilting angle <ω> (°) |
|---|---|---|---|---|---|---|---|---|
| 0.00 | Orthorhombic | 5.5326 | 5.5127 | 7.8254 | 238.674 | (O1) 1.9690(4) | 167.1(22) | 6.45 |
| | | | | | | (O2)$^{(1)}$ 1.880(10) | 153.0(6) | 13.50 |
| | | | | | | (O2)$^{(2)}$ 2.135(10) | 153.0(6) | 13.50 |
| 0.25 | Rhombohedral | 5.5165 | 5.5165 | 13.365 | 352.229 | 1.96550 | 162.760 | 8.62 |
| 0.50 | Rhombohedral | 5.4568 | 5.4568 | 13.422 | 346.116 | 1.93670 | 171.966 | 8.03 |
| 0.75 | Rhombohedral | 5.4549 | 5.4549 | 13.3179 | 343.195 | 1.92700 | 177.4(34) | 1.30 |
| 1.00 | Cubic | 3.85158 | 3.85158 | 3.85158 | 57.118 | 1.92558 | 180 | 0 |

Atomic positions for Pbnm are 4(c) (*x, y*, 1/4) for La, and 4(b) (1/2, 0, 0) for Fe/Ni, and 4(c) (*x, y*, 1/4) for O1, and 8(d) (*x, y, z*). The refined positions are 4(c) (0.0026, 0.02385, 1/4) for La, and 4(c) (0.040, 0.4959, 1/4) for O1 and 8(d) (-0.2916, 0.2582, 0.0542) for O2.

Atomic positions for R-3c are 6(a) (0, 0, 1/4) for La/Sr, and 6(b) (0, 0, 0) for Fe/Ni, and 8(e) (*x*, 0, 1/4) for O. The refined positions are 8(e) (0.447, 0, 1/4) for the sample x=0.25, and 8(e) (0.475, 0, 1/4) for the sample x=0.50, and 8(e) (0.508, 0, 0.25) for the sample x=0.75.

Atomic positions for Pm-3m are 6(a) (1/2, 1/2, 1/2) for Sr, 6(b) (0, 0, 0) for Fe/Ni, and 18(e) (1/2, 0, 0) for O.

With increasing Sr substitution, the symmetry of the system LSFN increases from orthorhombic to rhombohedral, and then to cubic. The changes in lattice parameters and unit cell volume depending on Sr doping concentration are illustrated in Figure 3(a) and Figure 3(b), respectively.

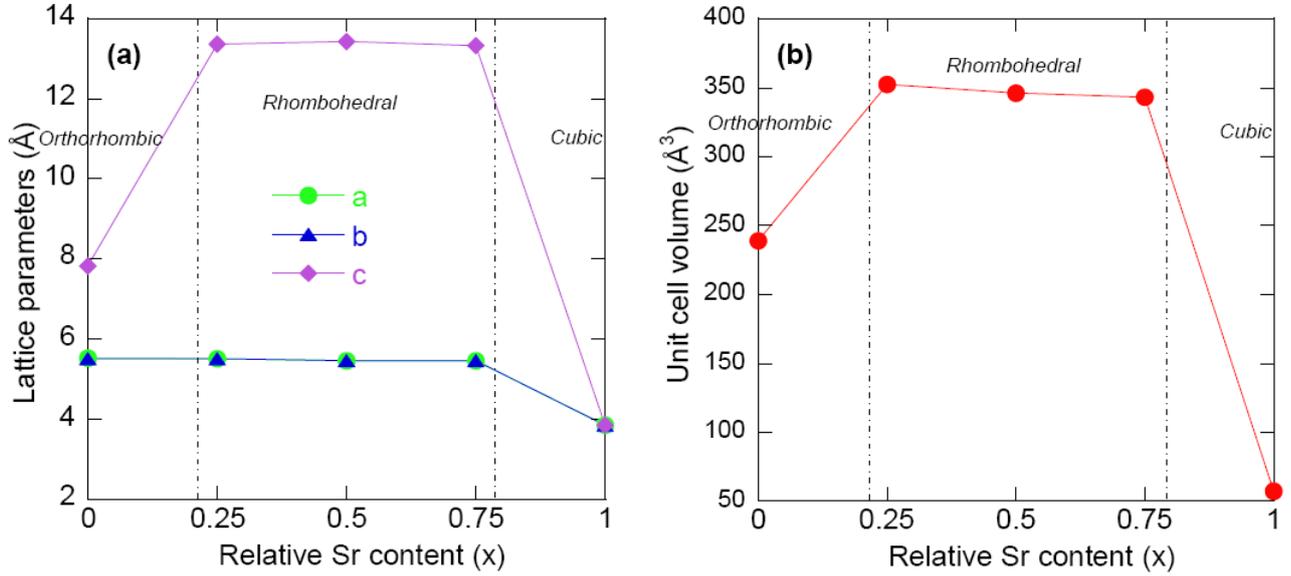

Figure 3. The changes in (a) lattice parameters and (b) unit cell volume of $La_{1-x}Sr_xFe_{0.75}Ni_{0.25}O_{3-\delta}$ obtained by Rietveld refinement.

Since LSFN with x=0 is in orthorhombic symmetry, two different oxygen positions (O1 and O2) exist with equal relative occupation of O1 (50%) and O2 (50%). There are two different distances between Fe/Ni–O2 which has also equal concentration (25%) and which are labeled $O2^{(1)}$ and $O2^{(2)}$. In the following, thus, we will consider the weighted arithmetic average distance <Fe/Ni–O> for comparison with the other samples:

$$\langle Fe/Ni - O \rangle = \frac{1}{2}\langle Fe/Ni - O \rangle + \frac{1}{4}\langle Fe/Ni - O^{(1)} \rangle + \frac{1}{4}\langle Fe/Ni - O^{(2)} \rangle$$

The superexchange angle θ, which is formed by the M–O–M (Fe/Ni–O–Fe/Ni) bridge in $ABO_3$ type perovskites, is 180° for cubic symmetry. The deviation of θ from 180° which is due to distortion in

[Fe/Ni]O$_6$ octahedra, is called tilting angle, and is directly related with the superexchange angle [24]: $\langle \omega \rangle = \frac{1}{2}(180 - \langle \theta \rangle)$.

In order to compare the superexchange and tilting angle of the samples the weighted angles are considered for x=0.

Goldschmidt tolerance factor t

$$t = (r_A + r_B)/\sqrt{2}(r_B + r_O)$$

allows us to estimate the degree of distortion in the perovskites. Since we have two different species on the A (La and Sr) and B (Fe and Ni) sites of the ABO$_3$ perovskites, we need to form the weighted average values [27] based on the ionic radii of the atoms calculated by Shannon [28] for different coordination numbers (CN) and spin state like low spin (LS) and high spin (HS). The ionic radii of the cations are 1.36 Å for La$^{3+}$$_{(CN:XII)}$; 1.42 Å for Sr$^{2+}$$_{(CN:XII)}$; 0.645 for Fe$^{3+}$$_{(CN:VI\ in\ HS)}$; 0.585 Å for Fe$^{3+}$$_{(CN:VI)}$; 0.56 Å for Ni$^{3+}$$_{(CN:VI\ in\ LS)}$; 0.60 Å for Ni$^{3+}$$_{(CN:VI\ in\ HS)}$. The ionic radii of O$^{2-}$$_{(CN:VI)}$ is 1.40 Å [28].

As long as the ions at the A-site (A' and A") and B-site (B' and B") are randomly distributed within the substituted perovskite A'$_{1-x}$A"$_x$B'$_{1-y}$B"$_y$O$_{3-\delta}$, the tolerance factor needs to be rewritten to be able to calculate the weighted average of valences and ionic radius:

$r_A = (1-x) \cdot La^{3+} + x \cdot Sr^{2+}$ and $r_B = (1-y) \cdot Fe^{\gamma} + y \cdot Ni^{3+}$

In our calculation, Ni is considered as Ni$^{3+}$ [21] for both low spin (LS) and high spin (HS) state because we could not identify the actual spin state unambiguously. However, we are confident about the average oxidation state of Fe ($\gamma$) as a result of the Ligand Field Multiplet Calculation (LFMC) for our Fe L$_{2,3}$ edge X-ray absorption spectra. The experimental data and their simulation with LFMC are compared for LaFeO$_3$ and 25% Ni doped LaFe$_{0.75}$Ni$_{0.25}$O$_3$ and shown in Figure 4. The Slater integrals used in the calculation for LaFeO$_3$ were scaled down to 70%, as already exercised by Abbate et al. [17] and for LaFe$_{0.75}$Ni$_{0.25}$O$_3$ to 50% of their atomic values in order to mimic covalence effects [21]. Thus it is clear that, once Ni is substituted to LaFeO$_3$, it decreases the d-d, p-d interaction and p-d exchange interaction due to the fact that the Fe 3d orbitals become broader and the overlap between Fe 3d and O 2p orbitals increases. Both, the pure LaFeO$_3$ and the Ni substituted LaFe$_{0.75}$Ni$_{0.25}$O$_3$ have to 100% Fe$^{3+}$ in the high spin $t_{2g}^3 e_g^2$ ($^6A_{1g}$) ground state. However, 25% Ni substitution increases the crystal field from 1.80 eV to 1.85 eV.

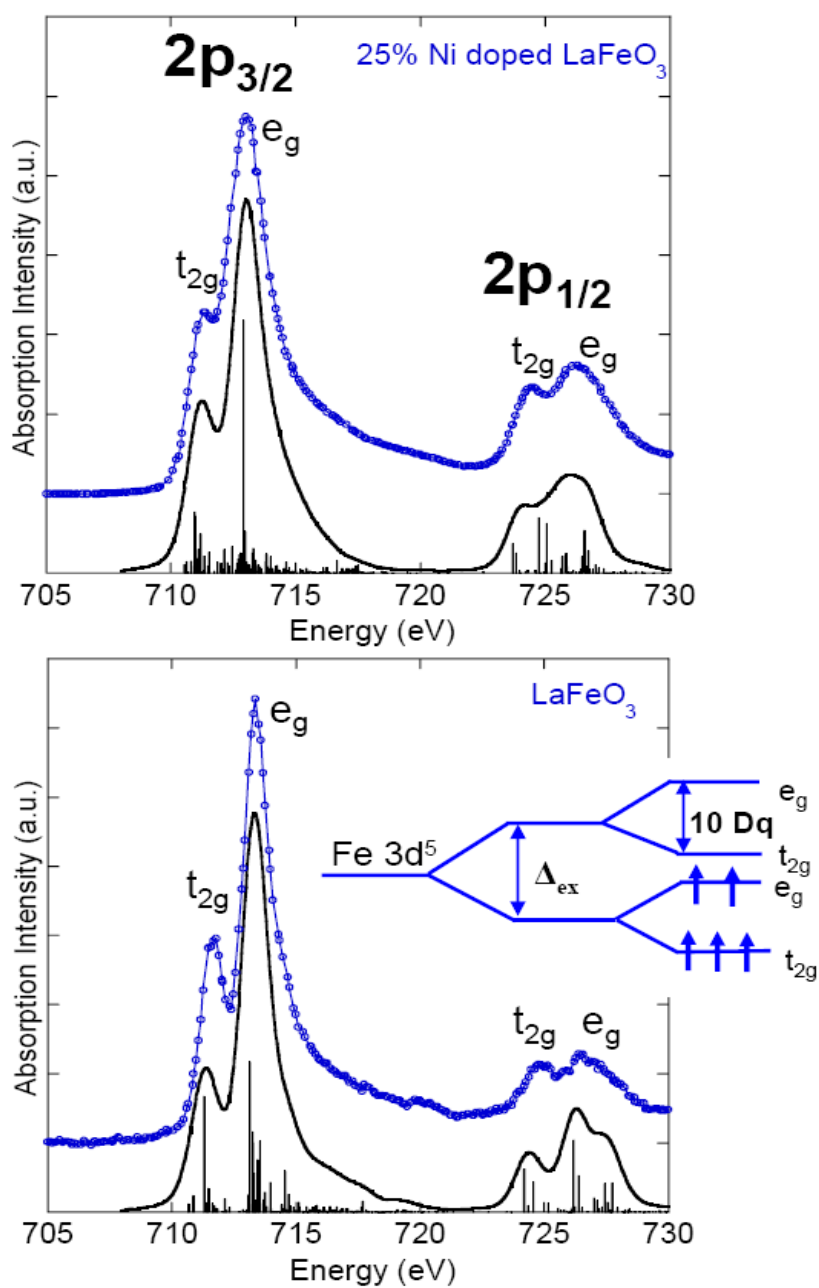

Figure 4. Comparison between experimental (blue line with open symbols) and simulated (black line, bottom) $Fe_{2,3}$ absorption spectra for $LaFeO_3$ and $LaFe_{0.75}Ni_{0.25}O_3$. Inset shows the electronic configuration of high spin Fe $3d^5$ in the ground state and modified after Abbate et. al [17]. Fe L edge split into $L_3$ ($2p_{3/2}$) and $L_2$ ($2p_{1/2}$) due to spin orbit coupling and additionally split into $t_{2g}$ and $e_g$ levels due to crystal field effect.

The comparison for x=0.25, 0.50, and 0.75 was published recently [21] and is rewritten for the reader in Table 2.

Tolerance factors are calculated for both $Ni^{3+}$ in LS and HS, and the results are listed in Table 2.

**Table 2.** Average oxidation state of Fe and the Tolerance factor (t) of LSFN with $Ni^{3+}$ both in low and high spin state.

| Sr content (x) | Oxidation state of Fe (γ) | t ($Ni^{3+}$ in LS) | t ($Ni^{3+}$ in HS) |
|---|---|---|---|
| 0.00 | 3.00 (100% $Fe^{3+}$ +0% $Fe^{4+}$) | 0.96436 | 0.95962 |
| 0.25 | 3.25 (75% $Fe^{3+}$ + 25% $Fe^{4+}$) | 0.97678 | 0.97196 |
| 0.50 | 3.50 (50% $Fe^{3+}$ +50% $Fe^{4+}$) | 0.98933 | 0.98442 |
| 0.75 | 3.95 (5% $Fe^{3+}$ +95% $Fe^{4+}$) | 1.00660 | 1.00150 |
| 1.00 | 4.00 (0% $Fe^{3+}$ + 100% $Fe^{4+}$) | 1.01490 | 1.00980 |

LS: Low spin state and HS: High spin state. Both $Fe^{3+}$ and $Fe^{4+}$ are in HS [21].

The calculated Goldschmidt tolerance factors for LSFN show a linear trend for both $Ni^{3+}$ spin state species (see Figure 5). The small differences in the ionic radii between $Ni^{3+}$ low spin and high spin reflect the differences in the tolerance factor. However, once La is replaced by Sr, the tolerance factor starts to increase linear which is an evidence for an increase in the symmetry of the system.

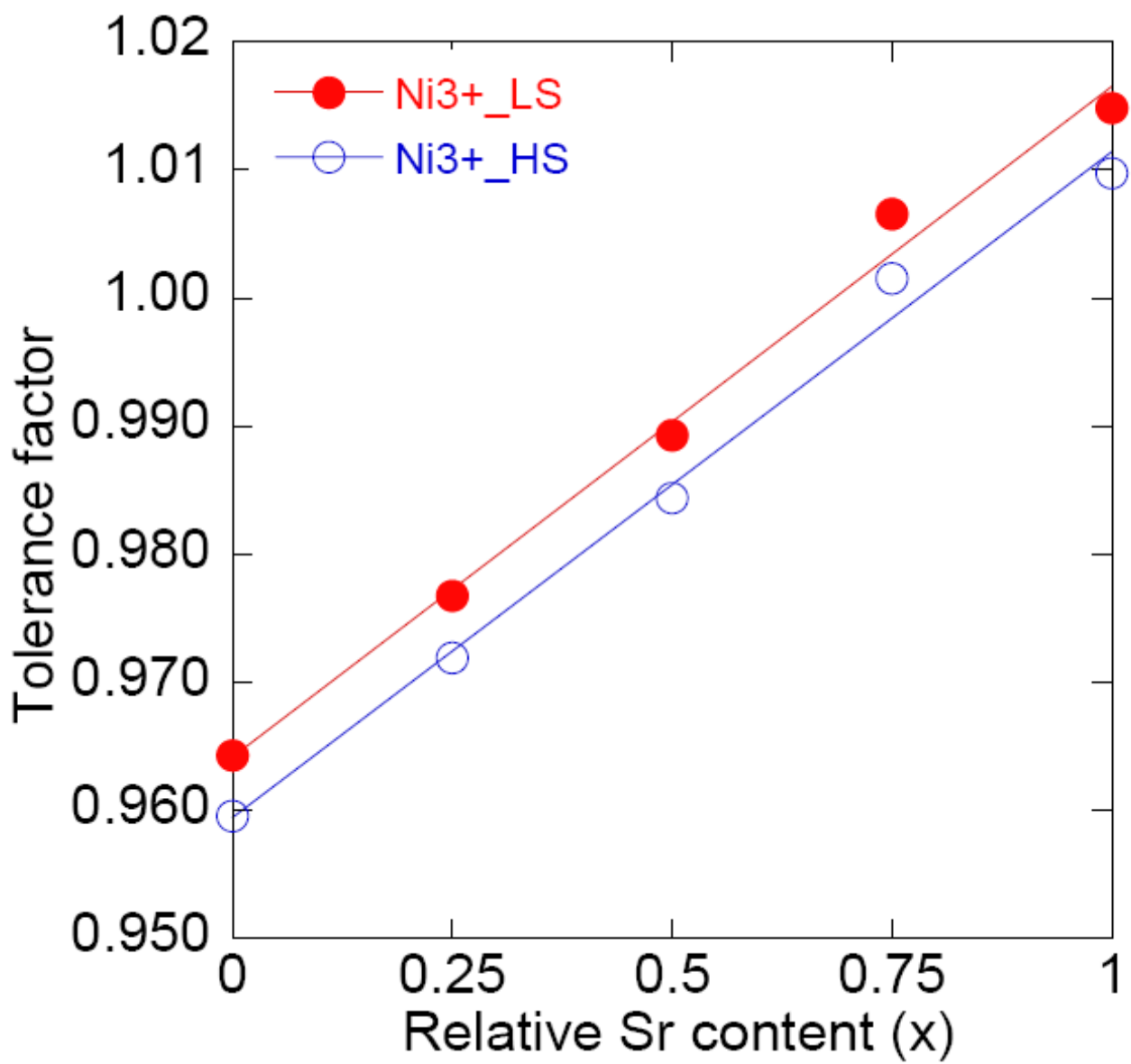

Figure 5. Tolerance factor of $La_{1-x}Sr_xFe_{0.75}Ni_{0.25}O_{3-\delta}$ depending on relative Sr content for $Ni^{3+}$ both in low spin and high spin.

The sample with x=0 has a smallest tolerance factor with $Ni^{3+}$ high (low) spin t=0.960 (0.964) for which Rietveld analysis shows it has orthorhombic symmetry. The samples with x=0.25, x=0.50, and x=0.75 have tolerance factor with $Ni^{3+}$ high (low) spin t=0.972 (t=0.977), t=0.984 (t=0.989), and t=1.002 (t=1.007) having rhombohedral symmetry. For x=1.0, the tolerance factor is t=1.015 (1.010), and the sample has cubic symmetry. For ideal cubic SrTiO3, t=1.00 and for orthorhombic GdFeO3, t=0.81 and it was mentioned that cubic phase occurs if 0.89<t<1.00 [29]. For our samples, the cubic phase occurs if t >1.0.

In perovskites, the A-site does not play a direct role on conductivity but has an indirect influence: the Coulomb potential barrier created by A-site around B-site affect the electron hole hopping process via $Fe^{3+}$–O–$Fe^{3+}$/$Fe^{4+}$ superexchange unit. Therefore, the A-site Coulomb potential ($Z_A/r_A$) where $Z_A$ and $r_A$ are the charge and ionic radius of A-site is calculated depending on Sr doping. For those samples which contain both La and Sr, the weighted charge and radius are used for the calculation. The results are illustrated in Figure 6, along with the Fe/Ni–O bond length.

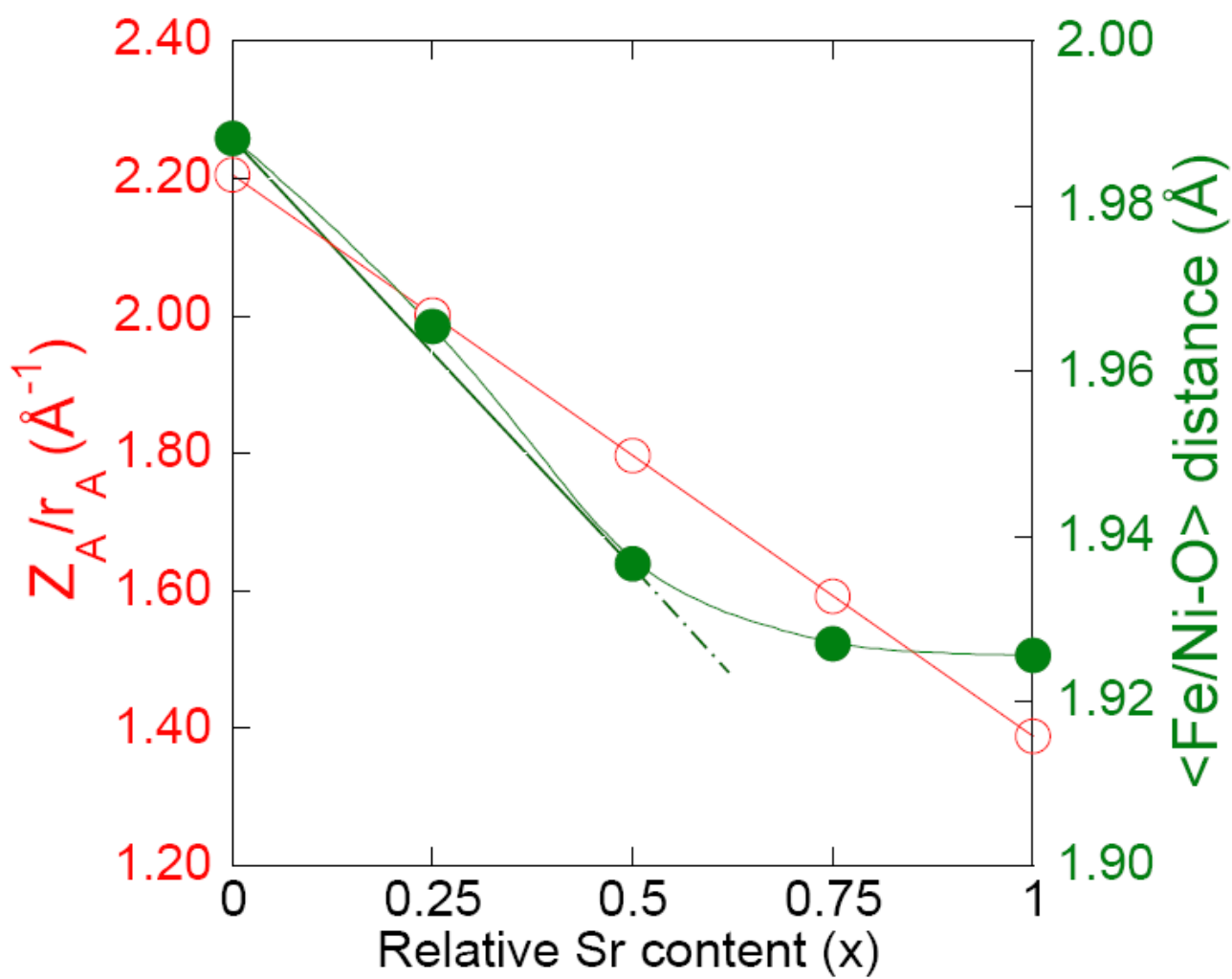

Figure 6. A-site Coulomb potential and Fe/Ni–O distance of $La_{1-x}Sr_xFe_{0.75}Ni_{0.25}O_{3-\delta}$.

As it is clearly shown in Figure 6, the A-site Coulomb potential decreases linear with increasing Sr content. This is in line with our expectation because La is 3+ and doped with lower charged Sr 2+

having larger ionic radius. The distance between Fe/Ni–O decreases linear up to 50% Sr doping and for $0.50 < x \leq 1.0$ it approaches a constant value of around 1.925 Å. Therefore, the variation of the Fe/Ni–O distance deviates from the linear behavior of $Z_A/r_A$ for $x > 0.50$, possibly because the system is not completely ionic but becomes more covalent.

**Electric conductivity**

The electrical conductivity of $La_{1-x}Sr_xFe_{0.75}Ni_{0.25}O_{3-\delta}$ was measured in the temperature range of 300 K $\leq T \leq$ 1273 K. Generally, this class of materials shows similar behavior depending on temperature; the conductivity increases with increasing temperature like a semiconductor ($d\rho/dT<0$) and then starts to decrease similar to metallic behavior ($d\rho/dT>0$). Furthermore, the transition temperature from semiconducting to metallic like behavior depends on the Sr doping. However, in this paper we would like to explain conductivity changes at 300 K depending on Sr content as shown in Figure 7. The conductivity at 300 K increases with increasing Sr content up to 50% and then starts to decrease. 25% and 50% of Sr doping in $LaFe_{0.75}Ni_{0.25}O_3$ increases the conductivity one and two orders of magnitude, respectively. However, for high Sr doping level (75%), the conductivity decreases by one order of magnitude as compared to 50% doping. The sample with x=1.0 was very brittle and could not be subjected to reliable conductivity measurements. However, test with multimeter tips showed a resistance in the MΩ range, suggesting that this material was an insulator.

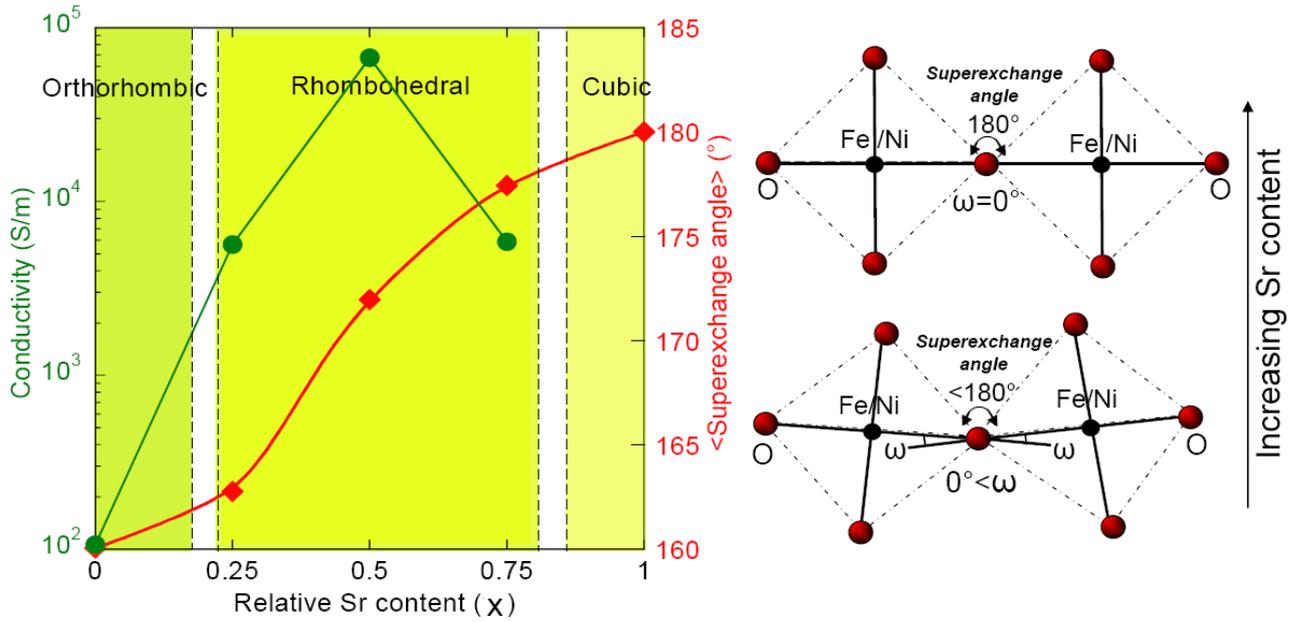

Figure 7. Electrical conductivity at 300 K and superexchange angle depending on Sr content. A fragment of the rhombohedral/orthorhombic symmetry after [24] with a schematic representation of the average superexchange angle, <Ni/Fe–O–Ni/Fe>, and the average tilting angle, <ω> which changes in to cubic symmetry with 180° superexchange angle and 0° tilting angle at high Sr doping.

**Discussions**

The electrical conductivity is caused by the conducting electron hole hopping process from $Fe^{3+}$ to $Fe^{4+}$ via oxygen bridge ($Fe^{3+}$–$O^{2-}$–$Fe^{4+}$) in LSF, and we concluded that charge transfer from oxygen to nickel (O–Ni) contributes additionally to electrical conductivity in LSFN [21]. Oxygen vacancies which interrupt the bridges between O and Fe/Ni are undesirable not only from the point of hopping but also from the point of charge transfer contributions. The oxygen vacancy concentration is increased with increasing Sr content in the samples with the formula of $La_{1-x}Sr_xFe_{0.75}Ni_{0.25}O_3$ [21].

In order to understand the electrical conductivity we compare the sample with x=0 (Ni doped LSF) with $LaFeO_3$. The sample with x=0 is a semiconductor while $LaFeO_3$ is a charge transfer insulator at 300 K. Replacing Fe by Ni causes a decrease in M–O distance resulting in an increase in O 2p bandwidth. Thus, the orbital overlap between M 3d and O 2p increases [30]. That might be a reason why the sample x=0 has charge transfer from O 2p to Ni 3d which makes it conducting, in compari-son to LaFeO$_3$ despite their similar crystallographic symmetry (orthorhombic) and electronic configuration 3d$^5$ with total spin S=5/2.

When we want to show the A-site Sr doping effect on the conductivity, it is necessary to discuss numerous parameters such as electronic interaction, charge transfer, symmetry including superexchange and/or tilting angle, A-site potential and M–O distance. Now that we have detailed quantitative information on the crystallographic structure and electronic structure of the LSFN samples, the electronic conductivity can be discussed and rationalized.

The sample with x=0 having antiferromagnetic Fe$^{3+}$–O$^{2-}$–Fe$^{3+}$ [31] and spin related superexchange interaction shows the least charge transfer. The charge transfer mainly depends on the overlap between O 2p and Ni 3d orbitals and is maximum when the <Ni–O–Ni> superexchange angle is 180° (cubic symmetry) or, in other words <ω>=0. Since this sample is in orthorhombic symmetry it has the largest deviation from the ideal cubic symmetry <Ni–O–Ni> superexchange angle (tilting angle) is around 160° (<ω>≈9.98°) with highest M–O distance due to strong A-site Coulomb potential and least charge transfer consequently.

The sample with x=0.25 has ferromagnetic Fe$^{3+}$–O$^{2-}$–Fe$^{4+}$ [31] hopping process with double exchange interaction where the conducting electrons jump from Fe$^{3+}$ toward Fe$^{4+}$ via the O bridge. As mentioned above, the A-site potential decreases when La$^{3+}$ is replaced by Sr$^{2+}$, resulting in a decrease in O–Fe/Ni distance, an increase in the O 2p bandwidth, and consequently an increase in mobility of the double exchange electrons and an increase in concentration of charge transferred electrons to Ni. Since the symmetry changes from orthorhombic to rhombohedral, the distortion is reduced (<ω>=8.62°) resulting in an increase in <Fe/Ni–O–Fe/Ni> superexchange angle of around 163°. This provides an additional increase in charge carrier mobility.

Similar rational applies to the sample with x=0.50: the A-site Coulomb potential decreases, the Fe/Ni–O distance gets shorter, keeps the symmetry rhombohedral with a lower distortion (<ω>=8.03°) and a larger <Fe/Ni–O–Fe/Ni> superexchange angle of around 172°, resulting in a increase in charge transfer. An important detail of the sample with x=0.50 is that the donor (Fe$^{3+}$) and acceptor ($Fe^{4+}$) concentration are equal, which gives additional increase in charge carrier mobility. Consequently, this sample shows the highest electrical conductivity.

In the case of high Sr doping (x=0.75), the electrical conductivity starts decreasing although the crystallographic symmetry comes closer to cubic with <Fe/Ni–O–Fe/Ni> superexchange angle around 177° and with weaker A-site potential consequently shorter Fe/Ni–O distance. That is why the sample has highest charge transfer. From this point the sample is expected to show the highest electrical conductivity. On the other hand, this sample allows for the $Fe^{3+}$–$O^{2-}$–$Fe^{4+}$ hopping process with double exchange interaction with 5% of donor and 95% acceptor resulting in a decrease in mobility of charge carriers. In addition to that, this sample has the highest oxygen vacancy concentration which breaks down the oxygen bridges between Fe/Ni–$O^{2-}$–Fe/Ni superexchange unit, and results in a decrease in double exchange mechanism. There are two mechanisms contributing electrical conductivity, charge transfer increases while the double exchange decreases in this sample. Since the electrical conductivity decreases in our experiment we can conclude that the decrease in double exchange is higher than the increase in charge transfer and thus over compensates this effect. This also brings another conclusion concerning to oxygen vacancies. The vacancies are most probably created around the Fe, and not around the Ni cation. If the oxygen vacancies were created around Ni, the charge transfer form O 2p to Ni 3d orbitals would decrease upon increasing Sr content.

In $La_{1-x}Sr_xFe_{0.75}Ni_{0.25}O_{3-\delta}$, the Fe concentration is three times higher than Ni. Therefore, for all samples it is expected that the $Fe^{3+}$–$O^{2-}$–$Fe^{4+}$ hopping process is dominant comparing to charge transfer. At this point it is worth to mention that we did not consider the hopping process across the $Ni^{3+}$–$O^{2-}$–$Ni^{3+}$ superexchange unit: provided that the atoms are randomly distributed in $La_{1-x}Sr_xFe_{0.75}Ni_{0.25}O_{3-\delta}$ to lower the total energy of the system, the probability of having Ni atoms next to each other is negligibly small.

**Conclusions**

The crystallographic phase transformation from orthorhombic to rhombohedral and then to cubic is observed with increasing Sr content in $La_{1-x}Sr_xFe_{0.75}Ni_{0.25}O_3$ at room temperature. An increase in calculated Goldschmidt tolerance factor shows that the symmetry of the system increases. The A-site potential is reduced by Sr substitution resulting in a decrease in distance between Fe/Ni–O distance leading an increase in overlap between O 2p and Fe/Ni 3d orbitals.

The electrical conductivity mechanisms of $La_{1-x}Sr_xFe_{0.75}Ni_{0.25}O_3$ depending on relative Sr content are discussed in terms of hopping process with super/double exchange interaction and charge transfer mechanism. The effect of crystallographic changes on the conductivity is discussed in detail. In the low doping region ($0 \leq x \leq 0.50$), both exchange interaction and charge transferred electron hole concentration increases with increasing Sr content, although in the high doping region (x=0.75), the exchange mechanism decreases while the charge transfer increases. In order to test such speculations, density functional theory calculations may prove helpful. Such calculations are in progress, with particular focus on the projected density states and its modification upon doping.

**Acknowledgement**

Financial support by the European Commission (MIRG # CT-2006-042095 and Real-SOFC # SES6-CT-2003-502612) and the Swiss National Science Foundation (SNF #200021-116688) are acknowledged. The ALS is supported by the Director, Office of Science, Office of Basic Energy Sciences, of the U.S. Department of Energy under Contract No. DE-AC02-05CH11231. The authors would like to thank to Dr. Lynne McCusker from ETH-Zurich for helpful discussions on crystallographic results.